\documentclass[prd,showpacs,preprintnumbers,amsmath,amssymb,superscriptaddress]{revtex4}
\usepackage{graphics}

\newcommand{\al}{\alpha}
\newcommand{\be}{\beta}
\newcommand{\ga}{\gamma}
\newcommand{\de}{\delta}

\newcommand{\pd}{\partial}

\newcommand{\dd}{{\rm d}}

\newcommand{\eps}{\varepsilon}

\begin{document}

\title{Multiple-scales analysis of cosmological perturbations in brane-worlds}
\author{Richard A. Battye}
\affiliation{Jodrell Bank Observatory, Department of Physics and
Astronomy, University of Manchester, Macclesfield, Cheshire SK11
9DL, UK}
\author{Andrew Mennim}
\affiliation{Department of Applied Mathematics and Theoretical
Physics, Centre for Mathematical Sciences, University of Cambridge,
Wilberforce Road, Cambridge CB3 OWA, UK}
\date{12 August 2004}
\preprint{DAMTP-2004-81}

\begin{abstract}
We present a new approximation method for solving the equations of
motion for cosmological tensor perturbations in a Randall--Sundrum
brane-world model of the type with one brane in a five-dimensional
anti-de Sitter spacetime.  This method avoids the problem of
coordinate singularities inherent in some methods.  At leading order,
the zero-mode solution replicates the evolution of perturbations in a
four-dimensional Friedmann--Robertson--Walker universe in the absence
of any tensor component to the matter perturbation on the brane.  At
next order, there is a mode-mixing effect, although, importantly, the
zero-mode does not source any other modes.
\end{abstract}

\maketitle

\section{Introduction}

The concept of brane-worlds is not a very new one~\cite{oldbraneworld}, but has risen
in popularity in the last five years.  
Central to this high level of interest has been the observation that the experimental
bounds on the inverse square behaviour of the gravitational force are
surprisingly large, with the current bound being about $10^{-4}\,$m~\cite{adelberger};
by contrast, the equivalent bound on the Coulomb law is about $10^{-18}\,$m.
In a seminal paper, Arkani-Hamed, Dimopoulos and Dvali~\cite{ADD}
observed that, if the Standard Model interactions were confined to a
four-dimensional subspace, the length-scale of extra dimensions could
be made as large as these bounds on Newton's law.
A further revelation came from the work of Randall and Sundrum~\cite{RS1,RS2}, who
demonstrated that the extra dimensions could be made infinite in
extent if one removed the assumption of a product geometry.

We will focus on the single-brane Randall--Sundrum (RS)
model~\cite{RS2}.  The cosmological background solution for this model
was found shortly after the model was suggested~\cite{RScos}; however,
the question of how perturbations evolve in such a background is still
unresolved.  Some progress has been made, for example, 
it is possible to analyse the evolution of perturbations 
perturbations during  a primordial de Sitter phase~\cite{LMSW,Rubakov}
of the  brane-world assuming an initial vacuum state. Our interest
here is in the a more general cosmological background dominated by
either radiation or matter.

Several approaches have been proposed to study this problem.  One is
to try to construct an effective theory in four dimensions and study
perturbations in that.  This is particularly useful for brane-world
models with a finitely sized extra dimension since it is then not
possible to excite Kaluza--Klein states below a certain energy;
however, it is less well motivated in the case of the single-brane model. 
In this method one ignores most of the higher-dimensional effects by
reducing the problem to a scalar-tensor theory, so even where it is
an accurate approximation it can never offer a ``smoking gun'' for
brane-world models by predicting a uniquely higher-dimensional effect.
Other approaches have been to try to solve the full five-dimensional problem,
either numerically~\cite{HKT} or analytically, using some
approximation scheme.  
Typically, these methods fall into two categories: brane-based and bulk-
The brane-based methods use a Gaussian normal (GN) coordinate system,
constructed from a congruence of geodesics normal to the brane, where
the brane is chosen to lie at a fixed value of one of the coordinates.
These GN coordinates make the boundary conditions at the brane simple
at the expense of rendering the equation of motion for the perturbations complicated.
The GN coordinates also suffer from a coordinate horizon, meaning that
they do not cover the whole spacetime.
In the case of a brane with the induced metric of de Sitter space, the
coordinate singularity coincides with the horizon, and, as mentioned
above, it is possible to solve the problem~\cite{LMSW,Rubakov}.

In the bulk-based approach one typically chooses a coordinate system that makes
manifest the symmetries of the bulk, resulting in a simple equation of
motion for the perturbations; the penalty for this is that the brane
is not at a fixed value of one of the coordinates and, consequently,
applying the boundary condition is complicated.
In the special case of a brane with a Minkowski induced metric, the
brane is at a fixed value of one of the coordinates and the
perturbation equations can be solved; indeed, this was done by Randall and
Sundrum did in their original paper~\cite{RS2}.
One major advantage of the bulk-based approach is that it does not
have any coordinate singularities.
The relation between the GN and bulk coordinates is presented in
Appendix~\ref{App:coord}.

Several methods for solving this problem have been suggested in the
literature, all of which have some limitations.
One possibility is to make an educated guess about the behaviour of the
bulk effects~\cite{guess} and treat these effects as a source
term in the usual equations for the evolution of perturbations in four
dimensions.  The obvious drawback is that the guess might not be
educated enough; furthermore, the perturbations on the brane can
source the bulk effects, meaning that a simple ansatz is unlikely to
give a realistic answer.
To go beyond this, many authors have suggested methods of
approximating the problem, for example, the near-brane
approximation~\cite{nearbrane} or the gradient expansion
method~\cite{gradexp}.  The main limitation with both of these
approaches is that they rely on the GN coordinates, which have a
coordinate horizon.

In this article, we present a new approximate method for solving the evolution
equations for tensor perturbations.  There are two stages to this: a
change of coordinates to a hybrid system --- a bulk-based system
modified to make the boundary condition tractable within a well-motivated 
approximation scheme --- and an adaptation of the method of multiple
scales~\cite{BenderOrszag,Hinch} to find approximate solutions to the
equations of motion.  Intuitively, this approximation works by
assuming that the boundary condition evolves slowly compared to the
time-scale of the solutions.
We will comment at the end on the consequences of the solutions we
obtain and on how we hope to extend this method to work for scalar
perturbations.

\section{The equations of motion}

We will start from the Poincar\'e coordinates, as advocated
in Ref.~\cite{nathalie}, where the metric is manifestly conformally related
to Minkowski space.
The line element has the form
\begin{equation}
\label{confminkds}
\dd s^2=\frac{l^2}{Z^2} \eta_{\al\be}\, \dd x^\al \dd x^\be
=\frac{l^2}{Z^2}\Big(-\dd T^2+\de_{ij}\,\dd x^i \dd x^j
+\dd Z^2\Big) \,,
\end{equation}
where $l$ is the AdS length-scale.  We have assumed spatial flatness
of the background cosmology here.
This coordinate system makes the five-dimensional linearized Einstein
equations simple, the price being that the brane has locus given by
$Z=l/a$, rather than being at a fixed value of one of the coordinates,
making the boundary condition much more difficult to impose.
If we write the perturbed line element as
\begin{equation}
\dd s^2=\frac{l^2}{Z^2}\Big( \eta_{\al\be}+h_{\al\be} \Big)
\dd x^\al \dd x^\be\,,
\end{equation}
and choose the transverse-traceless (TT) gauge,
following Ref.~\cite{nathalie}, where $h_{\al\be}$ satisfies  
$h_{\al Z}=0$, $\eta^{\al\be}h_{\al\be}=0$ and $\overline{\!\nabla}_\al
h^\al{}_{\!\be}=\bar{g}^\rho{}_{\!\ga}\,\bar{g}^\al{}_{\!\mu}\,
\bar{g}^\nu{}_{\!\be}\,\nabla_{\!\rho} h^\mu{}_{\!\nu}=0$, 
with $\bar{g}$ and $\overline{\nabla}$ representing, respectively, the
projected metric and derivative on the brane.
This setup has the advantage that the resulting perturbed
Einstein equations~\cite{nathalie} exactly soluble.
After making a Fourier transform $x^i \rightarrow k^i$
in the spatial directions parallel to the brane, the equation is
simply
\begin{equation}
\frac{\pd^2h}{\pd T^2}-\frac{\pd^2h}{\pd Z^2}+
\frac{3}{Z}\frac{\pd h}{\pd Z}+k^2 h=0\,,
\end{equation}
where the indices on $h$ have been dropped for convenience.
This equation can be solved by separation of variables to give mode functions
$h_m=\exp\big(\pm i \sqrt{k^2+m^2} T\big)\,Z^2{\cal B}_2\left(mZ\right)$,
where ${\cal B}_2$ is a Bessel function of order 2.  There will be five
different polarizations of the graviton in five-dimensions.
However, in this gauge the position of the brane is not on the same
locus as for the background but is displaced, an effect which has
been dubbed ``brane-bending'' in the literature.
This can be written as $\Delta n^\mu$, that is, a displacement of $\Delta$
from the background brane position along the direction of the unit normal
$n$, which is given by~\cite{nathalie}
\begin{equation}
n^\mu=\frac{1}{a}\left(\eps,0,0,0,-\sqrt{1+\eps^2}\right)\,,\qquad
\eps=\frac{l}{a^2}\frac{\dd a}{\dd \tau}=\frac{l}{a}\frac{\dd a}{\dd t}=lH\,,
\end{equation}
where $a$ is the scale factor on the brane, and $\tau$ and $t$ are,
respectively, conformal and proper time on the brane.
Combined with the five polarizations of the gravitational
perturbations in the bulk, this give six degrees of freedom,
corresponding to the scalar, vector and tensor degrees of freedom in
four-dimensional cosmology.
Since the perturbation of the position of the brane is a scalar degree of
freedom, we can ignore this effect when studying tensor perturbations
to linear order.  This approach should also apply to the bulk
perturbations which give rise to vector modes on the brane, although
we will not discuss these here.

The boundary condition at the brane is given by the Israel junction
condition~\cite{israel}, relating the jump in the normal derivative of
the metric to the matter supported on the brane.  (With reflection
symmetry imposed between the two sides of the brane, the jump is equal
to twice the value on one side.)  For tensor perturbations this is simply
\begin{equation}
\label{tensorjunction}
\left.{(n.\nabla)h^T_{ij}}\right|_{\text{brane}} = -\kappa\Sigma_{ij}^T\,,
\end{equation}
where the r.h.s.\ is the tensor part of the perturbation of the
brane stress-energy-momentum tensor.
For most of the history of the universe, there is no matter source for
the tensor modes so the r.h.s.\ of (\ref{tensorjunction}) will
reduce to zero.
Only when the CMB photons develop a quadrupole moment in the
late universe is this assumption no longer valid.

In summary, the problem we must solve for the evolution of the tensor
perturbations is
\begin{equation}
\frac{\pd^2h}{\pd T^2}-\frac{\pd^2h}{\pd Z^2}+
\frac{3}{Z}\frac{\pd h}{\pd Z}+k^2 h=0\,, \qquad\mbox{subject to}\qquad
\sqrt{1+\eps^2}\left.\frac{\pd h}{\pd Z}\right|_{Z=l/a}
=\eps\left.\frac{\pd h}{\pd T}\right|_{Z=l/a}\,,
\end{equation}
which, although seemingly innocuous, is extremely difficult to do exactly.
We will try to find approximate solutions based on the observation
that there are two independent dimensionless parameters in the
problem: $\eps=lH$ and $p=kl$, which are, respectively, the Hubble
parameter and the frequency of the perturbation measured in terms of
the AdS length-scale.
The presence of two scales means that the solutions should have time
dependence on two different scales --- much like an amplitude- or
frequency-modulated radio broadcast, which is a signal of a high 
frequency with its amplitude or frequency changing on a much longer
time-scale.
The method of multiple scales~\cite{BenderOrszag,Hinch} is
specifically designed to find approximate solutions in precisely these
situations.

\section{New coordinate system}

We will be constructing an approximation scheme valid when $\eps=lH$
is small.
If one naively applies the method of multiple scales as described in
section~\ref{Sec:multscale}, one obtains an incorrect answer.
This is because, to get the correct evolution at zeroth order, it is
necessary for the boundary condition to have no first-order term.
In order to make the method work, we devise a new set of coordinates
designed to remove the first-order term in the boundary condition.

From the change of coordinates in the appendix, we have
that, on the brane,
\begin{equation}
\frac{\dd\tau}{\dd T}=\sqrt{1+\eps^2}\,,
\end{equation}
where $\tau$ is conformal time for an observer on the brane moving
with the Hubble flow.  We can then evaluate
\begin{equation}
\frac{\dd a}{\dd T}=\frac{\dd\tau}{\dd T}\frac{\dd
  t}{\dd\tau}\frac{\dd a}{\dd t}=\frac{a^2}{l}\eps\sqrt{1+\eps^2}\,.
\end{equation}
If we also assume that the Friedmann law $H^2=8\pi G\rho/3$ holds to
sufficient accuracy, that is, we are late enough in the history of
the universe for the $\rho^2$ correction~\cite{RScos} to be ignorable,
then we can compute that
\begin{equation}
\frac{\dd H}{\dd t}=-4\pi G(\rho+p)=-\frac{3}{2}\,\Gamma\,H^2\,,
\end{equation}
for a cosmological fluid with polytropic index $\Gamma$ satisfying
$\rho+p=\Gamma\rho$.  This gives
\begin{equation}
\frac{\dd \eps}{\dd T}=-\frac{3\Gamma}{2l}\,a\,\eps^2\,\sqrt{1+\eps^3}\,.
\end{equation}

If we make the change of coordinates
\begin{equation}
\xi=\frac{Z}{l}\,,\qquad
\eta=\frac{T}{l}+\frac{a\eps Z^2}{2l^2}\,,
\label{newcoords}
\end{equation}
the equation of motion becomes
\begin{equation}
\frac{\pd^2 h}{\pd\eta^2}-\frac{\pd^2 h}{\pd\xi^2}+\frac{3}{\xi}\frac{\pd h}{\pd\xi}+
p^2h=2\eps a\left\{\xi\frac{\pd^2 h}{\pd\xi\pd\eta}-\frac{\pd h}{\pd\eta}\right\}
+{\cal O}(\eps^2)\,,
\end{equation}
where $p=kl$ is the dimensionless frequency of the mode.  The boundary condition is simply
\begin{equation}
\label{BCbrane}
\left.\frac{\pd h}{\pd \xi}\right|_{\xi=1/a}=
{\cal O}\big(\eps^3\big)\,.
\end{equation}
Note that the brane is still at $\xi=1/a$ in this new
coordinate system.

\section{Multiple-scales analysis}
\label{Sec:multscale}

For the purpose of multiple-scales analysis~\cite{BenderOrszag,Hinch},
we introduce two new time variables: ``fast'' time $f$, and ``slow''
time $s$, given by
\begin{equation}
f=\eta\,,\qquad s=\frac{1}{a}\,.
\end{equation}
Note that these are dimensionless (as is $\xi$).
The variables $f$ and $s$ are both, of course, functions of $\eta$ so the
definition is a technical construct to allow us simply to handle the
fact that the boundary condition changes with time on a scale related
to $s$.
Let us list our various time coordinates for clarity:

\begin{center}
\begin{tabular}{c@{\qquad}l}
$T$ & is the time coordinate in the static coordinates,\\
$\eta$ & is the time coordinate in the new coordinates,\\
$\tau$ & is conformal time on the brane,\\
$t$ & is proper time on the brane,\\
$f$ & is the ``fast'' time coordinate,\\
$s$ & is the ``slow'' time coordinate.
\end{tabular}
\end{center}

We assume that $\eps$ is small, which is valid except in the very 
early universe.
In the usual application of the method of multiple
scale~\cite{BenderOrszag,Hinch} the small parameter $\eps$ is a
constant; here it is a function of $T$.
We then write derivatives w.r.t.\ $\eta$ in terms of the fast and slow time
derivatives as
\begin{eqnarray}
\frac{\pd}{\pd\eta}&=&\frac{\pd}{\pd f}-
\eps\frac{\pd}{\pd s}+{\cal O}\big(\eps^2\big)\,,\\
\frac{\pd^2}{\pd\eta^2}&=&\frac{\pd^2}{\pd f^2}
-2\eps\frac{\pd^2}{\pd f\pd s}+{\cal O}\big(\eps^2\big)\,.
\end{eqnarray}
Expressed in terms of these fast and slow variables, the equation
of motion is
\begin{equation}
\label{eqmotion}
\frac{\pd^2h}{\pd f^2}-\frac{\pd^2h}{\pd\xi^2}+\frac{3}{\xi}
\frac{\pd h}{\pd\xi}+p^2 h=2\eps\left(\frac{\pd^2h}{\pd f\pd s}
-a\frac{\pd h}{\pd f}+a\xi\frac{\pd^2h}{\pd\xi\pd f}\right)
+{\cal O}\big(\eps^2\big)\,,
\end{equation}
subject to the boundary condition (\ref{BCbrane}) at $\xi=s$.

\section{Solutions}

We will use the separation of variables technique, where one tries to
find a solution of the form
\begin{equation}
h=\int \Psi_m(f,s,\eps)\,\Phi_m(\xi,s,\eps)\,\dd m\,.
\end{equation}
We will write $\Psi_m(f,s,\eps)$ and $\Phi_m(\xi,s,\eps)$ as series in $\eps$:
\begin{equation}
\Psi_m(f,s,\eps)\approx\Psi^{(0)}_m(f,s)+
\eps\,\Psi^{(1)}_m(\eta,s)+\dots\,,\qquad
\Phi_m(\xi,s,\eps)\approx\Phi^{(0)}_m(\xi,s)+
\eps\,\Phi^{(1)}_m(\xi,s)+\dots\,,
\end{equation}
which we will require to be asymptotic as $\eps \rightarrow 0$, and 
we will require that the $\Phi_m(\xi,s)$  satisfy the boundary
condition (\ref{BCbrane}) individually.

\subsection{Zeroth-order solution}

We will construct the solution order by order in $\eps$.
The equation of motion to zeroth order is
\begin{equation}
\frac{1}{\Psi^{(0)}}\frac{\pd^2\Psi^{(0)}}{\pd f^2}+p^2=
\frac{1}{\Phi^{(0)}}\frac{\pd^2\Phi^{(0)}}{\pd\xi^2}-\frac{3}{\xi}
\frac{1}{\Phi^{(0)}}\frac{\pd\Phi}{\pd\xi}\,,
\label{firsteqn}
\end{equation}
which is separable with solutions
\begin{eqnarray}
\label{Psizeroth}
\Psi_m^{(0)}(f,s)&=&\sum_{\pm}A_m^{\pm}(s)\exp\Big(\pm i\omega_m f\Big)\,,
\qquad \text{where}\qquad \omega_m=\sqrt{p^2+m^2}\,,\\
\Phi_m^{(0)}(\xi,s)&=&\frac{\pi}{2}m\xi^2\Big(Y_1(ms)\,J_2(m\xi)-J_1(ms)\,Y_2(m\xi)\Big)\,,
\label{Phisoln}
\end{eqnarray}
with $\Phi_m^{(0)}$ being chosen to satisfy the boundary condition
(\ref{BCbrane}) to zeroth order:
\begin{equation}
\left.\frac{\pd\Phi^{(0)}}{\pd\xi}\right|_{\xi=s}=0\,.
\end{equation}
The zero-mode solution $\Phi_0^{(0)}(\xi,s)=s$ is the limit of
(\ref{Phisoln}) as $m\rightarrow0$.
Of course, one has the freedom to redistribute the $s$ dependence
between $\Phi$ and $\Psi$ so as to leave the product unchanged. 
We can use the relation derived from the Wronksian of Bessel
functions, given in the Appendix B, to evaluate $\Phi_0^{(0)}(\xi,s)$ on the brane as
\begin{equation}
\Phi_m^{(0)}(s,s)=s=\frac{1}{a(\tau)}\,.
\end{equation}
Note that what would be constants, fixed by the initial conditions, in
a standard solution of (\ref{firsteqn}) are now functions of the slow
time variable, $s$, and it is these functions which encode the crucial
evolution of the amplitude.

\subsection{First-order solution}

The function $A_m(s)$ in expression (\ref{Psizeroth}) is undetermined by the
zeroth-order calculation; in order to calculated it, we must consider
the first order equation and construct the so-called secularity
condition which forces the solution to asymptotic.
The order $\eps$ terms in the equation of motion (\ref{eqmotion}) give us
\begin{eqnarray}
\label{firstorder}
&&\int\dd m\,\Phi^{(0)}_m\left(\frac{\pd^2\Psi^{(1)}_m}{\pd f^2}
+\omega_m^2\Psi^{(1)}_m\right)
-\int\dd m\,\Psi^{(0)}_m\left(\frac{\pd^2\Phi^{(1)}_m}{\pd z^2}
-\frac{3}{z}\frac{\pd\Phi^{(1)}_m}{\pd z}+m^2\Phi^{(1)}_m\right)\nonumber\\
&&\qquad\qquad{}=2\int\dd m\left(\frac{\pd^2\Psi^{(0)}_m}{\pd f\pd s}\Phi^{(0)}_m
+\frac{\pd\Psi^{(0)}_m}{\pd f}\frac{\pd\Phi^{(0)}_m}{\pd s}
-\frac{1}{s}\frac{\pd\Psi^{(0)}_m}{\pd f}\Phi^{(0)}_m
+\frac{\xi}{s}\frac{\pd\Psi^{(0)}_m}{\pd f}\frac{\pd\Phi^{(0)}_m}{\pd\xi}\right)\,,
\end{eqnarray}
and the boundary condition to order $\eps$ is the same as at zeroth order
\begin{equation}
\label{firstorderBC}
\left.\frac{\pd\Phi^{(1)}_m}{\pd\xi}\right|_{\xi=s}=0\,.
\end{equation}
We still have some freedom in constructing the solution, which we
shall use to impose
\begin{equation}
\frac{\pd^2\Phi^{(1)}_m}{\pd\xi^2}-\frac{3}{\xi}\frac{\pd\Phi^{(1)}_m}{\pd\xi}+m^2\Phi^{(1)}_m=0\,,
\end{equation}
so that $\Phi^{(1)}_m$ satisfies the same equation as $\Phi^{(0)}_m$.
Given that the boundary condition for $\Phi^{(1)}_m$ is also the same
as $\Phi^{(0)}_m$, it is actually be possible to impose the stronger
condition $\Phi^{(1)}_m\equiv0$ without any loss of generality.
We can then use the orthogonality relation,
\begin{equation}
\label{ortho}
\int_{s}^\infty\frac{\dd\xi}{\xi^3}\,\Phi^{(0)}_m\,\Phi^{(0)}_n
=\frac{\pi^2}{4}m\Big(\big(J_1(ms)\big)^2+\big(Y_1(ms)\big)^2\Big)\,\de(m-n)\,,
\end{equation}
to write the equation of motion as an equation for $\Phi^{(1)}_m$,
giving
\begin{eqnarray}
\frac{1}{2}\left[\frac{\pd^2\Psi^{(1)}_m}{\pd f^2}+\omega^2_m\Psi^{(1)}_m\right]&=&
\frac{\pd^2\Psi^{(0)}_m}{\pd f\pd s}-\frac{1}{s}\frac{\pd\Psi^{(0)}_m}{\pd f}
+\frac{4}{m\pi^2}\Big(J_1(ms)^2+Y_1(ms)^2\Big)^{-1}\times\nonumber\\
&&\qquad\left[\int \dd n\,\frac{\pd\Psi^{(0)}_n}{\pd f}\,\int\,\frac{\dd\xi}{\xi^3}
\Phi^{(0)}_m \frac{\pd\Phi^{(0)}_n}{\pd s}
+\frac{1}{s}\int \dd n\,\frac{\pd\Psi^{(0)}_n}{\pd f}\,\int\,\frac{\dd\xi}{\xi^2}
\Phi^{(0)}_m \frac{\pd\Phi^{(0)}_n}{\pd\xi}
\right]\,.
\end{eqnarray}
The calculation of the various integrals involving Bessel functions is
explained in Appendix B.  We can write the equation of motion for $\Psi^{(1)}_m$ as
\begin{eqnarray}
\label{Psi1eq}
\frac{1}{2}\left[\frac{\pd^2\Psi^{(1)}_m}{\pd f^2}+\omega^2_m\Psi^{(1)}_m\right]&=&
\frac{\pd^2\Psi^{(0)}_m}{\pd f\pd s}+
\frac{\dd}{\dd s}\log\sqrt{\big(J_1(ms)\big)^2+\big(Y_1(ms)\big)^2}
\frac{\pd\Psi^{(0)}_m}{\pd f}+\int \dd n \:{\cal M}_{mn} \frac{\pd\Psi^{(0)}_n}{\pd f}\,,
\end{eqnarray}
for $m \ne 0$, and 
\begin{equation}
\label{Psi1zeromodeeq}
\frac{1}{2}\left[\frac{\pd^2\Psi^{(1)}_0}{\pd f^2}+\omega^2_m\Psi^{(1)}_0\right]=
\frac{\pd^2\Psi^{(0)}_0}{\pd f\pd s}+\int \dd n \:{\cal M}_{0n} \frac{\pd\Psi^{(0)}_n}{\pd f}\,,
\end{equation}
for the zero-mode.
The trace-free matrix ${\cal M}_{mn}$ is defined as
\begin{equation}
{\cal M}_{mn}=\frac{4}{\pi^2 m}\Big(J_1(ms)^2+Y_1(ms)^2\Big)^{-1}\left\{
\int\,\frac{\dd\xi}{\xi^3}\Phi^{(0)}_m \frac{\pd\Phi^{(0)}_n}{\pd s}
+\frac{1}{s}\int\,\frac{\dd\xi}{\xi^2}\Phi^{(0)}_m \frac{\pd\Phi^{(0)}_n}{\pd\xi}\right\}\,,
\end{equation}
for $m \ne n$ and is calculated in Appendix B; these terms do not
contribute to the secularity condition.

Substituting in the solution (\ref{Psizeroth}) for $\Psi^{(0)}_0$,
we get terms on the r.h.s.\ of (\ref{Psi1zeromodeeq}) for the form
\begin{equation}
\sum_{\pm}\pm ip e^{\pm i p f}\:\frac{\dd A_0^\pm}{\dd s}\,.
\end{equation}
This is a forcing term in equation (\ref{Psi1eq}) which, as a function of the fast time $f$,
has the same frequency as the natural frequency of the l.h.s. and so
will cause a resonance.  If $\eps$ decays as $1/\eta$ or slower, this
is forbidden since the $h^{(1)}$ term would not be of asymptotic order.
So we can impose the \emph{secularity condition} that $A_0^\pm(s)$ must
satisfy
\begin{equation}
\frac{\dd A_0^\pm}{\dd s}=0\qquad\Rightarrow\qquad
A^\pm_0=C^\pm\,,
\end{equation}
where the $C^\pm$ are constants independent of $s$.

For the massive modes $(m\ne 0)$,
 the requirement that (\ref{Psi1eq}) be free of
forcing terms gives the secularity condition
\begin{equation}
\frac{\dd}{\dd s}\left[A_m^\pm(s)\sqrt{\big(J_1(ms)\big)^2+\big(Y_1(ms)\big)^2}
\right]=0\qquad\Rightarrow\qquad
A_m^\pm(s)=C_m^\pm\Bigg(\big(J_1(ms)\big)^2+\big(Y_1(ms)\big)^2\Bigg)^{-1/2}\,,
\end{equation}
where, again, the $C_m^\pm$ are constants independent of $s$.
We will choose $C_m^\pm \propto m$ so that the limit as 
$m \rightarrow 0$ is a constant.
With the secularity condition imposed, equation of motion for
$\Psi^{(1)}_m$ is then
\begin{equation}
\frac{\pd^2\Psi^{(1)}_m}{\pd f^2}+\omega_m^2\Psi^{(1)}_m=
2i\int \dd n\,{\cal M}_{mn}\,A^{\pm}_m\,\omega_n\,
e^{\pm i\omega_n f}\,,
\end{equation}
which will have solution
\begin{equation}
\label{Psi1}
\Psi^{\pm(1)}_m=B_m^{\pm}(s)\,e^{\pm i\omega_m f}+2i
\sum_\pm \int \dd n\,
\frac{A^{\pm}_m\,\omega_n}{\omega_m^2-\omega_n^2}\,
{\cal M}_{mn}\,e^{\pm i\omega_n f}\,.
\end{equation}
In order to determine $B_m(s)$, one would have to consider its
secularity condition, coming from the second-order equations.

\section{Mode-Mixing and the Zero-Mode}

A great advantage of the method of multiple scales is that the
leading-order solution includes the slow-time variation derived
from the secularity condition so is usually a very good
approximation.
The modes do not influence each other at leading order, and our
calculation gives the zero-mode as
\begin{equation}
h_0(\eta,\xi)=\frac{1}{a} \, e^{\pm ip\eta}\,,
\end{equation}
and the massive modes as
\begin{equation}
h_m(\eta,\xi)=\frac{\pi m^2 \xi^2
\Big(Y_1(m/a)\,J_2(m\xi)-J_1(m/a)\,Y_2(m\xi)\Big)}
{2\sqrt{\big(J_1(m/a)\big)^2+\big(Y_1(m/a)\big)^2}}
\:e^{\pm i\omega_m\eta}\,,
\end{equation}
which takes the value
\begin{equation}
h_m(\eta,1/a)=\frac{m/a}
{\sqrt{\big(J_1(m/a)\big)^2+\big(Y_1(m/a)\big)^2}}
\:e^{\pm i\omega_m\eta}\,,
\end{equation}
on the brane.

By re-writing the equation of motion in a different way, we can see
more directly that the zero-mode to leading order gives the same answer as
four-dimensional cosmology. 
By treating one of the order $\eps$ terms in the equation of motion as
a zeroth order term, one gets
\begin{equation}
\frac{\pd^2h}{\pd f^2}+2\eps a\frac{\pd h}{\pd f}-\frac{\pd^2h}{\pd\xi^2}+\frac{3}{\xi}
\frac{\pd h}{\pd\xi}+p^2 h=2\eps\left(\frac{\pd^2h}{\pd f\pd s}
+a\xi\frac{\pd^2h}{\pd\xi\pd f}\right)+{\cal O}\big(\eps^2\big)\,.
\end{equation}
Setting the r.h.s.\ to zero for the zeroth-order calculation and
performing the separation of variables in the same way as before gives
the time evolution equation
\begin{equation}
\frac{\pd^2\Psi}{\pd f^2}+\frac{2}{a}\frac{\dd a}{\dd f}
\frac{\pd\Psi}{\pd f}+k^2\Psi=0
\end{equation}
for the time component of the zero-mode, which is exactly the same as
the equation in four dimensions.  Not surprisingly, the secularity
condition at first order reveals no slow-time dependence for the zero-mode.
So, by slightly dishonest accountancy with the $\eps$ terms, one can
make the time dependence of the zero-mode look exactly like the
four-dimensional mode.

At first order in $\eps$, there is a modification in the phase due to
the $\eps$-dependent term in (\ref{newcoords}).  Evaluated on the
brane, we have
\begin{equation}
\eta=\frac{\tau}{l}+\frac{\eps}{2a}+{\cal O}\big(\eps^2\big)\,,
\end{equation}
so the zeroth-order part of the solution on the brane can we written
\begin{equation}
\frac{m/a}
{\sqrt{\big(J_1(m/a)\big)^2+\big(Y_1(m/a)\big)^2}}
\:\exp\left\{\pm
i\omega_m\left(\frac{\tau}{l}+\frac{\eps}{2l}\right)\right\}
\end{equation}
for the massive mode, or
\begin{equation}
\frac{1}{a} \, \exp\left\{\pm ik\left(\tau+\frac{\eps l}{2a}\right)\right\}
\end{equation}
for the zero-mode.
This is in addition to the contribution of the $\Psi^{(1)}\Phi^{(0)}$ term.

The first-order equation in $\eps$ (\ref{Psi1}) has an integral term
on the r.h.s., which describes how a mode is sourced by the other modes. 
The matrix ${\cal M}_{mn}$, which is a function of $s$, is calculated
in Appendix B.  Of particular note is that the zero-mode cannot
source any other modes, that is, ${\cal M}_{0n}=0$.
Massive modes, however, can decay into other modes.  Again, this is an
effect which occurs at first order in $\eps$.  The effects of
mode-mixing would be very small later in the history of the universe
so one would expect all the interesting signatures coming from
mode-mixing to arise when $\eps$ is not too small, that is, when the
Hubble radius is only one or two orders of magnitude larger than the
AdS length-scale.  Of course, our approximation scheme breaks down as
$\eps$ approaches unity.

From studies of the evolution of perturbations during a de Sitter era
in the history of the universe~\cite{LMSW,Rubakov} we know that
production of massive modes is heavily suppressed.
So it is reasonable to take the zero-mode as an approximation for the
spectrum of perturbations after the end of inflation.
If the AdS length-scale, $l$, is sufficiently small that our
approximation method is valid at the end of inflation, then our
analysis shows that no massive perturbations will be sourced during
the subsequent evolution, except possibly at ${\cal O}\big(\eps^2\big)$.

\section{Discussion and Conclusions}
\label{conc}

The approximation method we have used has several advantages over the
alternatives.
Firstly, our coordinates are related to the Poincar\'e coordinates
advocated by~\cite{nathalie} by a non-singular transformation and so,
unlike GN coordinates do not have a coordinate singularity.
This means that the answers we get are valid right up to the horizon,
allowing one meaningfully to impose initial conditions on a Cauchy
surface or a boundary condition of no incoming radiation at the Cauchy
horizon.
Secondly, we have not had to specify the evolution of the scale
factor, $a$, so the method can be applied quite generally in different
eras of the universe and at transitions between such eras;
by contrast, when using a GN coordinate system the time evolution has
to be specified in order to find approximate solutions and, even then,
the equations are only tractable in certain cases~\cite{nearbrane}.
Finally, it should be simple to extend the method to other bulk
geometries where the perturbation spectrum is known.

In particular, it is worth comparing our method to the gradient
expansion method.  The gradient expansion method works by expanding
the solution in the bulk in orders of the derivatives normal to the
brane.  The small parameter is exactly the same one as used here.
The two methods are quite similar, the main difference
being that the method presented here uses a coordinate system free
from coordinate horizons.
Our multiple-scales method can also allow for the presence of very
massive modes, although intuition tells us not expect such modes to be
important at low energies.

An important case is when the initial state for our calculation is the
zero-mode solution, because this is what one expects from a de Sitter
phase in the history of the universe~\cite{LMSW,Rubakov}.
Our analysis shows that the evolution of the solution will be the same
as for perturbations in a four-dimensional FRW universe and that the
massive modes are not sourced.
This is somewhat obvious from physical intuition if these modes are
viewed as Kaluza--Klein particles because one would expect particles to
decay into less massive states, not more massive ones.
The picture would be very different if there were sources on the brane:
massive modes could be sourced by the brane-based matter.
For simplicity, and because it is a reasonable approximation to make
for the tensor part of the spectrum, we have assumed that there is no
matter source on the brane to source the bulk metric perturbations.
If one were to consider scalar perturbations, the brane sources could
not be neglected in the same way. 

Our calculation applies strictly to the tensor component of the metric
perturbations on the brane since they are gauge invariant, but it
should be simple to apply it to the vector modes.
Tensor modes are gauge-invariant, but for vector modes there is the
added complication of gauge choice;  the gauge we have used here is different
from the gauges usually used to study four-dimensional cosmological
perturbations, for example, the synchronous gauge.
It should also be possible to extend this method to allow for a tensor
matter source, such as the photon quadrupole which develops after
last scattering.
In addition, we hope, in future work, to include the effect of
perturbing the brane position, so as to apply the method to study
scalar perturbations.

\begin{acknowledgments}
RAB is supported by PPARC and AM is supported by Emmanuel
College, Cambridge.
\end{acknowledgments}

\appendix

\section{Coordinate change}
\label{App:coord}

We consider the coordinate change between Gaussian Normal (GN) coordinates
and the conformally flat bulk coordinates.  The metrics are given
respectively by
\begin{equation}
\dd s^2=-N^2\dd\tau^2+\dd\zeta^2+A^2\de_{ij}\dd x^i \dd x^j=
\frac{l^2}{Z^2}\left(-\dd T^2+\dd Z^2+\de_{ij}\dd x^i \dd x^j\right)\,.
\end{equation}
We use a trick~\cite{CarterUzan} to simplify the transformation by
noting that $-\dd T$ is a Killing vector, so that its inner product
with the normal vector $\pd/\pd\zeta$ is independent of $\zeta$ along
the geodesics in the $\zeta$ direction, and hence we can write
\begin{equation}
\frac{\pd T}{\pd\zeta}=E(\tau)\,.
\end{equation}
The usual formula for coordinate change gives
\begin{eqnarray}
-N^2&=&-\frac{l^2}{Z^2}\left(\frac{\pd T}{\pd\tau}\right)^2+\frac{l^2}{Z^2}
\left(\frac{\pd Z}{\pd\tau}\right)^2\,,\\
0&=&-\frac{\pd T}{\pd\tau}\frac{\pd T}{\pd\zeta}+
\frac{\pd Z}{\pd\tau}\frac{\pd Z}{\pd\zeta}\,,\\
1&=&-\frac{l^2}{Z^2}\left(\frac{\pd T}{\pd\zeta}\right)^2+\frac{l^2}{Z^2}
\left(\frac{\pd Z}{\pd\zeta}\right)^2\,.
\end{eqnarray}
Solving these simultaneously, we find
\begin{equation}
\left(\frac{\pd Z}{\pd\zeta}\right)^2=N^2E^2\,.
\end{equation}
On the brane, $N=A=a(\tau)$ and $Z=l/a$ so
\begin{equation}
\left.\frac{\pd Z}{\pd\tau}\right|_{\zeta=0}=
-\frac{l}{a^2}\frac{\dd a}{\dd\tau}=-\eps\,.
\end{equation}
Thus, $E(\tau)=-\eps/a(\tau)$, where the sign can be determined by
considering the components of the normal vector in the $T,Z$
coordinates.
So, the Jacobian matrix and its inverse are
\begin{equation}
\renewcommand{\arraystretch}{1.5}
\left( \begin{array}{cc} \frac{\pd T}{\pd\tau}&\frac{\pd Z}{\pd\tau}\\
\frac{\pd T}{\pd\zeta}&\frac{\pd Z}{\pd\zeta}
\end{array} \right)=
\left( \begin{array}{cc} N\sqrt{\frac{Z^2}{l^2}+\frac{\eps^2}{a^2}}&
-\frac{\eps N}{a}\\ -\frac{\eps}{a}&\sqrt{\frac{Z^2}{l^2}+\frac{\eps^2}{a^2}}
\end{array} \right)\,,
\qquad\quad
\renewcommand{\arraystretch}{1.5}
\left( \begin{array}{cc} \frac{\pd\tau}{\pd T}&\frac{\pd\zeta}{\pd T}\\
\frac{\pd\tau}{\pd Z}&\frac{\pd\zeta}{\pd Z}
\end{array} \right)=
\left( \begin{array}{cc} \frac{A}{N}\sqrt{1+\frac{\eps^2 A^2}{a^2}}&
\frac{\eps A^2}{a}\\ -\frac{\eps A}{aN}&A\sqrt{1+\frac{\eps^2 A^2}{a^2}}
\end{array} \right)\,.
\end{equation}

\section{Integrals of products of Bessel functions and Orthogonality relations}
\label{App:integrals}

\subsection{Standard results}

There is a standard formula for Bessel functions (see, for example,
Ref.~\cite{Watson}) which tells us that 
\begin{equation}
\label{Besselformula}
\int x B_\nu(mx) \tilde{B}_\nu(nx) \,\dd x = \frac{x}{m^2-n^2}
\Big( m B_{\nu+1}(mx) \tilde{B}_\nu(nx) -
n B_\nu(mx) \tilde{B}_{\nu+1}(nx) \Big) \,,
\end{equation}
where $B_\nu$ and $\tilde{B}_\nu$ are any two cylinder functions of
order $\nu$ and $m \ne n$.
We will apply this to two specific cylinder function defined by
\begin{eqnarray}
C^{(m)}_\nu(x)&=&mY_1(ms)J_\nu(x)-mJ_1(ms)Y_\nu(x)\,,\\
D^{(m)}_\nu(x)&=&Y_2(ms)J_\nu(x)-J_2(ms)Y_\nu(x)\,,
\end{eqnarray}
with $J$ and $Y$ being Bessel functions of the first and second kinds
respectively.
Cylinder functions satisfy the recurrence relations~\cite{Watson}
\begin{equation}
\label{Besselrecurrence}
zB_{\nu-1}(z)+zB_{\nu+1}(z)=2\nu B_\nu(z)\qquad\text{and}\qquad
B_{\nu-1}(z)-B_{\nu+1}(z)=2\frac{\dd B_\nu(z)}{\dd z}\,,
\end{equation}
from which it can be seen that
\begin{equation}
\frac{\pd}{\pd s}C^{(m)}_\nu(mx)=\frac{1}{s}C^{(m)}_\nu(mx)
-m^2 D^{(m)}_\nu(mx)\,,
\end{equation}
where the l.h.s. is also a cylinder function of order $\nu$.
The Wronskian of Bessel functions of the first and second
kinds is~\cite{Watson}
\begin{equation}
W_{[J_\nu,Y_\nu]}(z)=J'_\nu(z) Y_\nu(z) - Y'_\nu(z) J_\nu(z)
=Y_{\nu-1}(z) J_\nu(z) - J_{\nu-1}(z) Y_\nu(z)=\frac{2}{\pi z}\,,
\end{equation}
which, in conjunction with the recurrence relations, allows us to
evaluate the following
\begin{equation}
\begin{array}{l@{\qquad}l}
C^{(m)}_2(ms)=\frac{2}{\pi s}\,,&\frac{\pd}{\pd s}C^{(m)}_2(ms)=\frac{2}{\pi s^2}\,,\\
C^{(m)}_3(ms)=\frac{8}{\pi ms^2}\,,&\frac{\pd}{\pd s}C^{(m)}_3(ms)=\frac{8}{\pi ms^3}-\frac{2m}{\pi s}\,,\\
D^{(m)}_2(ms)=0\,,&D^{(m)}_3(ms)=\frac{2}{\pi m s}\,,
\end{array}
\end{equation}
which will be of use to use later on.

We will also make use of the asymptotic expansions for large argument~\cite{Watson}
\begin{equation}
J_\nu(x)\sim\sqrt{\frac{2}{\pi x}} \, \cos \left(x-\frac{\pi}{2}\nu-\frac{\pi}{4}\right)\,,\qquad
Y_\nu(x)\sim\sqrt{\frac{2}{\pi x}} \, \sin \left(x-\frac{\pi}{2}\nu-\frac{\pi}{4}\right)\,,
\end{equation}
the integral~\cite{Watson}
\begin{equation}
\label{mzeroint}
\int x^{-p+1} B_p(x) \dd x=-x^{-p+1}B_{p-1}(x)\,,
\end{equation}
and the orthogonality relation~\cite{Jackson}
\begin{equation}
\label{Jacksonexpr}
\int^\infty_0 x J_\nu(mx) J_\nu(nx)=\frac{1}{m} \de (m-n)\,,
\end{equation}
when $\nu$ is a whole number.

\subsection{Orthogonality relation}

From (\ref{Besselformula}) it is clear that, for $m \ne n$
\begin{equation}
\int^\infty_{x_0} x \, C^{(m)}_\nu(mx) \, C^{(n)}_\nu(nx) \,\dd x=0\,.
\end{equation}
because the derivatives of the $C^{(m)}$ functions with respect to $z$
vanish on the boundary $z=s$.
To determine the coefficient of the delta function in the
orthogonality relation, we note that the integral is dominated by the
behaviour near infinity, so we can consider instead the leading order
asymptotic expression
\begin{equation}
C^{(m)}_2(mx) \sim \sqrt{\frac{2m}{\pi x}} \Big(Y_1(ms)^2+J_1(ms)^2\Big)^{1/2}
\sin\big(mx+\text{phase angle}\big)\,,
\end{equation}
valid when $x$ is large.
So the constant of proportionality can be deduced to be
\begin{equation}
\int^\infty_{x_0} x C^{(m)}_\nu(mx)C^{(n)}_\nu(nx) \,\dd x
=m\Big(Y_1(ml)^2+J_1(ml)^2\Big)\,\de(m-n)\,,
\end{equation}
by comparison to the equivalent asymptotic expression for eq.~(\ref{Jacksonexpr}). 

This argument breaks down in the case of the zero-mode, which must be
evaluated separately.  The inner product of the zero-mode with itself
is simply
\begin{equation}
\int_s^\infty \frac{\dd\xi}{\xi^3} \Phi^{(0)}_0\Phi^{(0)}_0=
\int_s^\infty \frac{\dd\xi}{\xi^3} s^2 =\frac{1}{2}\,,
\end{equation}
while (\ref{mzeroint}) gives us
\begin{equation}
\int_s^{\infty}\frac{\dd\xi}{\xi^3} \Phi^{(0)}_m\Phi^{(0)}_0=0\,.
\end{equation}

\subsection{Resonant terms}

The integrals in the first order equation of motion (\ref{firstorder})
which need to be calculated to obtain the secularity condition are
\begin{eqnarray}
\label{firstresterm}
&&\int_s^\infty\frac{\dd\xi}{\xi^3}\,\Phi^{(0)}_m
\,\frac{\pd\Phi^{(0)}_m}{\pd s}\,,\\
\label{secondresterm}
&&\int_{s}^\infty\frac{\dd\xi}{\xi^2}\,\Phi^{(0)}_m\,\frac{\pd\Phi^{(0)}_m}{\pd\xi}\,,
\end{eqnarray}
To evaluate these when $m=0$ is simple, because
$\Phi^{(0)}_0=s$, and we can easily see that
\begin{equation}
\int_s^\infty \frac{\dd\xi}{\xi^3}\,\Phi^{(0)}_0\,\frac{\pd\Phi^{(0)}_0}{\pd s}
=\frac{1}{2s}\,,\qquad
\int_{s}^\infty\frac{\dd\xi}{\xi^2}\,\Phi^{(0)}_0\,\frac{\pd\Phi^{(0)}_0}{\pd\xi}=0\,.
\end{equation}

To evaluate (\ref{firstresterm}) when $m \ne 0$, we can differentiate the orthogonality relation
(\ref{ortho}) w.r.t.\ $s$ to get
\begin{equation}
\int_s^\infty\frac{\dd\xi}{\xi^3}\,\left(
\Phi^{(0)}_m\,\frac{\pd\Phi^{(0)}_n}{\pd s}+
\frac{\pd\Phi^{(0)}_m}{\pd s}\,\Phi^{(0)}_m\right)
=\frac{\pi^2}{4}m\frac{\dd}{\dd s}
\Bigg(\big(J_1(ms)\big)^2+\big(Y_1(ms)\big)^2\Bigg)\de(m-n)
+\frac{1}{s^3}\Phi^{(0)}_m(s,s)\Phi^{(0)}_n(s,s)\,.
\end{equation}
and note that, for $m=n$, the last term is ignorable compared to the delta
function term.
Evaluating (\ref{secondresterm}) when $m \ne 0$, can be done integrating by parts to get
\begin{equation}
\int_{s}^\infty\frac{\dd\xi}{\xi^2}\,\Phi^{(0)}_m\,\frac{\pd\Phi^{(0)}_m}{\pd\xi}=
\int_{s}^\infty\frac{\dd\xi}{\xi^3}\,\Phi^{(0)}_m\,\Phi^{(0)}_m
+\frac{1}{2}\left[\xi^{-2}\Phi^{(0)}_m\,\Phi^{(0)}_m\right]^\infty_s\,,
\end{equation}
which is just given by the orthogonality relations since, again, the second term
is insignificant compared to the first.

\subsection{Mode-mixing terms}

The mode-mixing, that is, the effect of modes as source terms for other
modes, arises from the terms involving the following integrals
\begin{eqnarray}
\label{firstmixterm}
&&\int_s^\infty\frac{\dd\xi}{\xi^3}\,\Phi^{(0)}_m
\,\frac{\pd\Phi^{(0)}_n}{\pd s}\,,\\
\label{secondmixterm}
&&\int_s^\infty\frac{\dd\xi}{\xi^2}\,\Phi^{(0)}_m
\,\frac{\pd\Phi^{(0)}_n}{\pd\xi}\,,
\end{eqnarray}
when $m \ne n$.  These integrals are simple to calculate when either
$m=0$ or $n=0$, using $\Phi^{(0)}_0=s$.  For (\ref{firstmixterm}) we
calculate that
\begin{equation}
\int_s^\infty \frac{\dd\xi}{\xi^3}\,\Phi^{(0)}_m\,\frac{\pd\Phi^{(0)}_0}{\pd s}=0\,,\qquad
\int_s^\infty \frac{\dd\xi}{\xi^3}\,\Phi^{(0)}_0\,\frac{\pd\Phi^{(0)}_n}{\pd s}=\frac{1}{s}\,,
\end{equation}
while (\ref{secondmixterm}) is evaluated to be
\begin{equation}
\int_{s}^\infty\frac{\dd\xi}{\xi^2}\,\Phi^{(0)}_m\,\frac{\pd\Phi^{(0)}_0}{\pd\xi}=0\,,\qquad
\int_{s}^\infty\frac{\dd\xi}{\xi^2}\,\Phi^{(0)}_0\,\frac{\pd\Phi^{(0)}_n}{\pd\xi}=
\frac{1}{n}C^{(n)}_2(ns)=\frac{2}{\pi n s}
\end{equation}
by applying the relation (\ref{mzeroint}).

When $m$ and $n$ are both non-zero, we can evaluate
(\ref{firstmixterm}) using the formula (\ref{Besselformula}) to give
\begin{equation}
\int_s^\infty\frac{\dd\xi}{\xi^3}\,\Phi^{(0)}_m
\,\frac{\pd\Phi^{(0)}_n}{\pd s}=\frac{\pi^2}{4}\left[
\frac{\xi}{m^2-n^2}\left(mC^{(m)}_3(m\xi)\frac{\pd}{\pd s}C^{(n)}_2(n\xi)-
nC^{(m)}_2(m\xi)\frac{\pd}{\pd s}C^{(n)}_3(n\xi)\right)\right]_{s}^\infty\,.
\end{equation}
The recurrence formulae (\ref{Besselrecurrence}) allow us to write
\begin{eqnarray}
C^{(m)}_\mu(m\xi)&=&\left[\frac{4}{s}Y_2(ms)-mY_3(ms)\right]J_\mu(m\xi)-
\left[\frac{4}{s}J_2(ms)-mJ_3(ms)\right]Y_\mu(m\xi)\,,\\
\frac{\pd}{\pd s}C^{(n)}_\nu(n\xi)&=&\left[\left(\frac{4}{s^2}-n^2\right)
Y_2(ns)-\frac{n}{s}Y_3(ns)\right]J_\nu(n\xi)-\left[\left(\frac{4}{s^2}
-n^2\right)J_2(ns)-\frac{n}{s}J_3(ns)\right]Y_\nu(n\xi)\,,
\end{eqnarray}
whence, we see that
\begin{equation}
\int_s^\infty\frac{\dd\xi}{\xi^3}\,\Phi^{(0)}_m
\,\frac{\pd\Phi^{(0)}_n}{\pd s}=-\frac{4}{(m+n)ms^3}
-\frac{mn}{(m^2-n^2)s}\,.
\end{equation}

To calculate (\ref{secondmixterm}) we first note that
\begin{equation}
\frac{\pd}{\pd\xi}\Big(\xi^2J_2(n\xi)\Big)=n\xi^2J_1(n\xi)
=\frac{\xi}{n}\frac{\pd}{\pd n}\Big(n^2J_2(n\xi)\Big)
\end{equation}
and similarly for $Y_2$.
Now
\begin{eqnarray}
\frac{\pd\Phi^{(0)}_n}{\pd\xi}&=&\frac{\pi}{2}n\Bigg[
Y_1(ns)\frac{\pd}{\pd\xi}\Big(\xi^2J_2(n\xi)\Big)-
J_1(ns)\frac{\pd}{\pd\xi}\Big(\xi^2Y_2(n\xi)\Big)\Bigg]\,,\nonumber\\
&=&\frac{\pi}{2}\xi\Bigg[
Y_1(ns)\frac{\pd}{\pd n}\Big(n^2J_2(n\xi)\Big)-
J_1(ns)\frac{\pd}{\pd n}\Big(n^2Y_2(n\xi)\Big)\Bigg]\,,
\end{eqnarray}
and
\begin{eqnarray}
\frac{\pd\Phi^{(0)}_n}{\pd n}&=&\frac{\pi}{2}\frac{\xi^2}{n}\Bigg[
Y_1(ns)\frac{\pd}{\pd n}\Big(n^2J_2(n\xi)\Big)-
J_1(ns)\frac{\pd}{\pd n}\Big(n^2Y_2(n\xi)\Big)\Bigg]\nonumber\\
&&{}+\frac{\pi}{2}\xi^2\Bigg[n^2\frac{\pd}{\pd n}\Bigg(\frac{Y_1(ns)}{n}\Bigg)
J_2(n\xi)-n^2\frac{\pd}{\pd n}\Bigg(\frac{J_1(ns)}{n}\Bigg)Y_2(n\xi)\Bigg]\nonumber\\
&=&\frac{\xi}{n}\frac{\pd\Phi^{(0)}_n}{\pd\xi}
+\frac{\pi}{2}\xi^2\Bigg[n^2\frac{\pd}{\pd n}\Bigg(\frac{Y_1(ns)}{n}\Bigg)
J_2(n\xi)-n^2\frac{\pd}{\pd n}\Bigg(\frac{J_1(ns)}{n}\Bigg)Y_2(n\xi)\Bigg]\nonumber\\
&=&\frac{\xi}{n}\frac{\pd\Phi^{(0)}_n}{\pd\xi}
-\frac{\pi}{2}ns\xi^2\Big(Y_2(ns)J_2(n\xi)-J_2(ns)Y_2(n\xi)\Big)
\end{eqnarray}
so that
\begin{equation}
\frac{\pd\Phi^{(0)}_n}{\pd\xi}=\frac{n}{\xi}\frac{\pd\Phi^{(0)}_n}{\pd n}
+\frac{\pi}{2}n^2s\,\xi\Big(Y_2(ns)J_2(n\xi)-J_2(ns)Y_2(n\xi)\Big)
\end{equation}
and we can write
\begin{equation}
\int_{s}^\infty\frac{\dd\xi}{\xi^2}\,\Phi^{(0)}_m\,\frac{\pd\Phi^{(0)}_n}{\pd\xi}
=n\frac{\pd}{\pd n}\int_{s}^\infty\frac{\dd\xi}{\xi^3}\,
\Phi^{(0)}_m\,\Phi^{(0)}_n+\frac{\pi}{2}n^2s\int_{s}^\infty\frac{\dd\xi}{\xi}
\,\Phi^{(0)}_mD^{(n)}_2(n\xi)\,.
\end{equation}
The first integral on the r.h.s.\ is zero when $m \ne n$; the second integral is
\begin{equation}
\frac{\pi^2}{4}n^2s\int_{s}^\infty \xi \,C^{(m)}_2(m\xi)\,D^{(n)}_2(n\xi)\dd \xi
=\frac{\pi^2}{4}\int_{s}^\infty \!\xi C^{(m)}_2(m\xi)\,C^{(n)}_2(n\xi)\,\dd \xi
-\frac{\pi^2}{4}s\int_{s}^\infty \!\xi C^{(m)}_2(m\xi)
\frac{\pd}{\pd s}C^{(n)}_2(n\xi)\,\dd \xi
\end{equation}
the first term of which is zero for $m \ne n$ by the orthogonality relation, the second having
already been calculated above.

\end{document}